\documentclass[12pt]{iopart}

\usepackage{iopams, epsfig}  

\begin{document}

\title[The DVCS Measurement at HERA]{The DVCS Measurement at HERA} 
 
\author{Ewelina \L obodzi\'{n}ska   
} 
 
\address{ DESY, Hamburg, Germany \\ 
Institute of Nuclear Physics, Krak\'{o}w, Poland}

\begin{abstract} 
The recent results of the studies of Deeply Virtual Compton Scattering (DVCS) events at HERA are presented. 
The possibility offered by this process to gain information about skewed parton distributions (SPD) is emphasized. 
 
\end{abstract} 
 
 
 
 
\section{Introduction} 
\subsection{Motivations} 
\begin{figure}[h] 
\begin{center} 
\mbox{ 
\epsfig{file=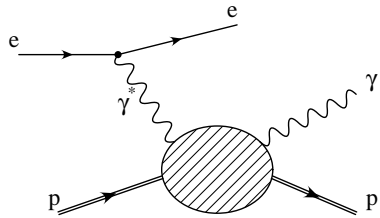, 
height=3.cm, 
clip=} 
} 
\end{center} 
\vspace*{-0.2cm}
\caption{\label{dvcs} The DVCS process} 
 
\end{figure}

The Deeply Virtual Compton Scattering process (DVCS) -- shown diagrammatically in Figure~\ref{dvcs} -  is a 
diffractive production of a real photon in deeply inelastic scattering. 
 
The apparent simplicity of this process makes it a new and powerful tool to study various aspects of QCD in 
the field of diffraction. 
However, the main interest comes from the fact that DVCS gives a comparatively clean access to new 
parton distributions, i.e. the skewed parton distributions (SPD) \cite{spd}. 
SPD are the generalization of the usual parton 
distributions to the case where the momentum transfer to the proton is non-zero. 
\begin{figure}[h] 
\begin{center} 
\mbox{ 
\epsfig{file=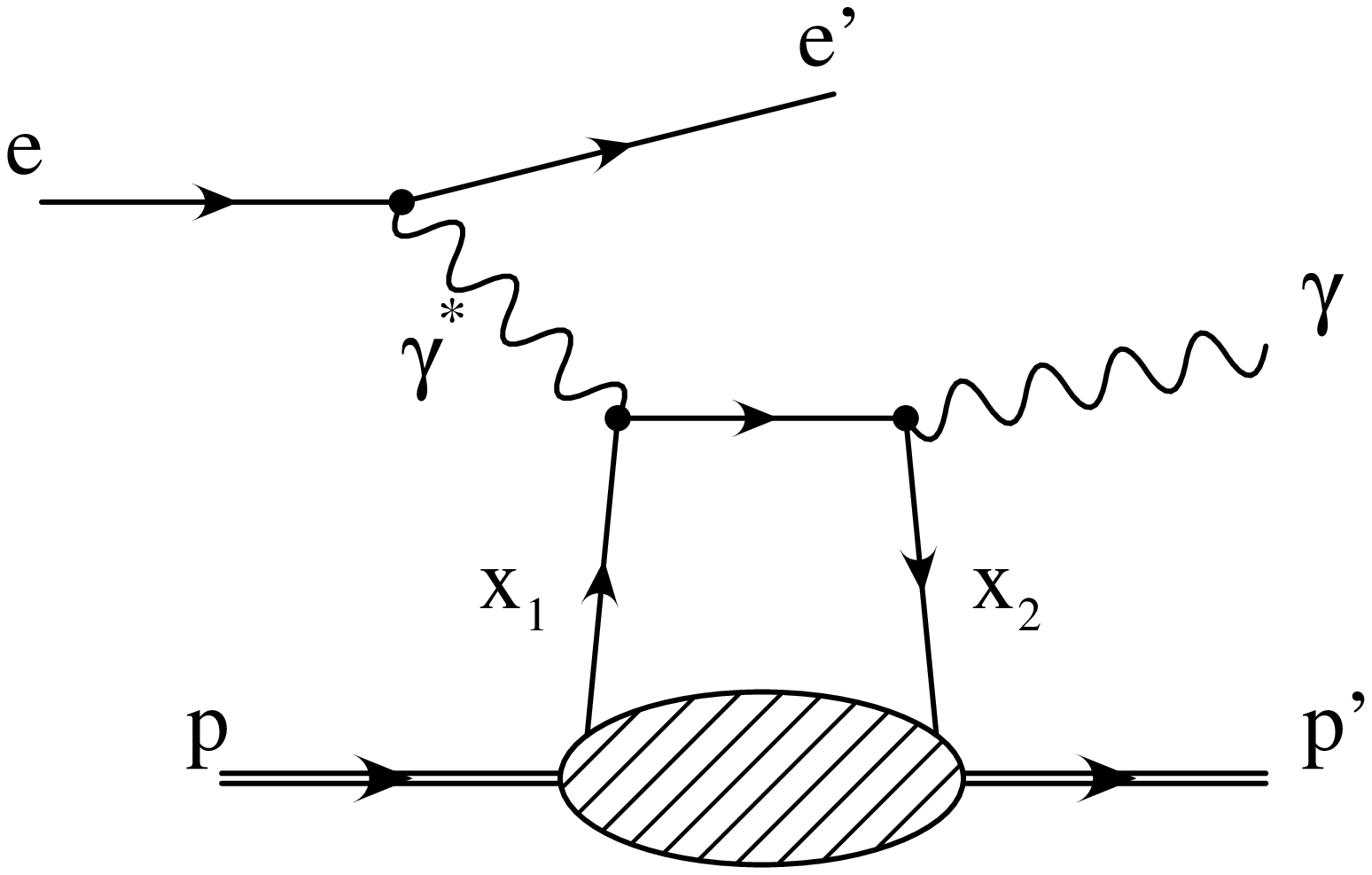, 
height=4.cm, 
width=5cm, 
bbllx=60pt,bblly=255pt, 
bburx=500pt,bbury=560pt, 
clip=} 
\epsfig{file=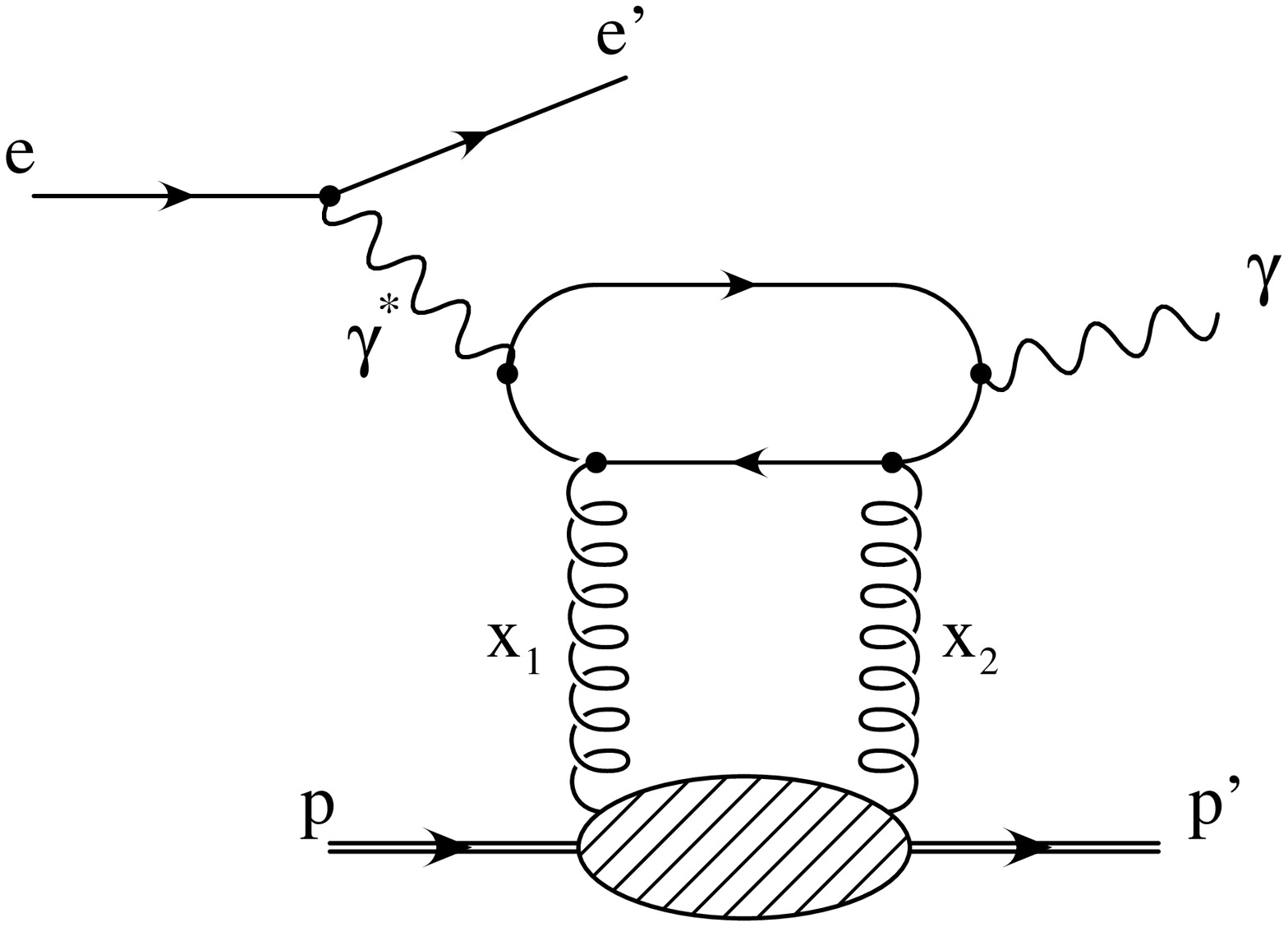, 
height=4.cm, 
width=8cm, 
bbllx=-115pt,bblly=180pt, 
bburx=550pt,bbury=560pt, 
clip=} 
} 
\end{center}
\vspace*{-0.5cm} 
\caption{\label{spd} The two dominant QCD diagrams for DVCS.\\ 
$x_1$ and $x_2$ are the fractions of the incoming proton momentum carried by the partons. 
} 
\end{figure} 
 This is illustrated in 
Figure~\ref{spd}, where two dominant QCD diagrams for DVCS are shown. 
The parton with the 
fraction $ x_{1}$ of the incoming proton momentum 
leaves the proton and returns to it with the momentum fraction $x_{2}$. It can be noticed 
that in order to bring the outgoing photon onto its mass shell, the fractions of the 
 momentum carried by the partons must be unequal (actually, $x_{1}$ - $x_{2}$ = $x_{B}$, where $x_{B}$ is the 
Bjorken variable \cite{spd,factorization}). 
 DVCS is the most desirable process for extracting SPD because: 
\begin{itemize}   
\parskip=0pt
   \parsep=0pt
   \itemsep=0pt 
\item it interferes with Bethe-Heitler process - as discussed in more detail in the next subsection - and SPD appear linearly in the interference term,
 \item it has a proven QCD factorization 
formula, so there is a reliable theoretical basis for extracting parton distributions \cite{factorization}, 
\item it is least suppressed in $Q^{2}$ among all known exclusive diffractive processes, so it is accessible over a broad range of $Q^{2}$,
\item the theoretical uncertainty connected with the process is minimized because the real final state photon is an 
 elementary particle, so there is no need for the meson wave function as in the case of 
vector mesons. 
\end{itemize} 
During the last years, the DVCS process gained a considerable theoretical interest  
\cite{factorization} -- \cite{new_theory}, mainly in the context of SPD. Quite recently, first observations and 
measurements have been reported \cite{zeus}--\cite{clas}. 
 
\subsection{Theoretical discussion} 
\label{discussion} 
\begin{figure}[h] 
\begin{center} 
\mbox{ 
\epsfig{file=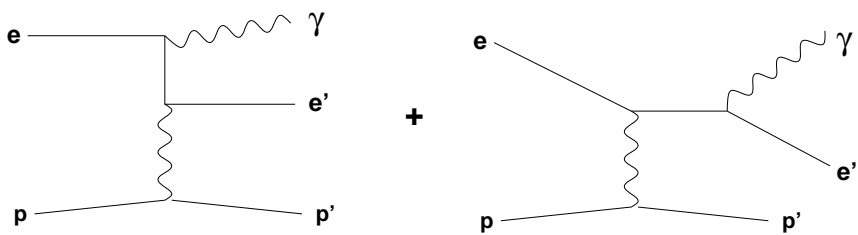, 
height=3.cm, 
clip=} 
} 
\end{center}
\vspace*{-0.2cm} 
\caption{\label{bh} The Bethe-Heitler process} 
\end{figure}

The reaction  
\begin{equation} 
 e^+ + p \rightarrow e^+ + p + \gamma  
\label{reaction} 
\end{equation} 
 receives contributions from both DVCS, whose 
origins lie in the strong interaction processes (Figure~\ref{spd}), and the purely electromagnetic 
Bethe-Heitler (BH) process (Figure~\ref{bh}). 
The total cross section for exclusive photon production described by the reaction (\ref{reaction}) consists of 
three parts:  
\begin{equation} 
d\sigma ^{total} = d\sigma^{DVCS} + d\sigma^{BH} + d\sigma^{interf} 
\label{cross} 
\end{equation} 
where 
 $ d\sigma^{DVCS} $ is the pure DVCS cross section, 
 $d\sigma^{BH} $ describes the BH contribution and 
$ d\sigma^{interf} $ corresponds to interference between the BH and DVCS processes. 
The BH process is well known as it depends only on QED calculations and the proton elastic form factors, so its 
cross section is precisely determined.  
It is also known that the interference term for the unpolarized positron beam is, in the leading twist approximation, proportional to $\cos\phi$, \ 
where $ \phi $ stands 
for the difference in angles of the positron and the proton in the transverse scattering plane.  
Therefore, the 
interference term vanishes for  all analyzes averaging over the full azimuthal angle of final states particles. 
In particular, in such a case the DVCS cross section can be extracted by 
subtracting the BH cross section from the total one. 
On the other hand, the measurement of the interference term gives the best access to SPD. The experimental 
observable to obtain SPD is 
the azimuthal angle asymmetry: 
\begin{equation} 
\mathcal{A} = \frac{\int_{-\pi /2}^{\pi /2}d\phi (d\sigma^{total} - 
d\sigma^{BH})  - 
\int_{\pi /2}^{3\pi /2}d\phi(d\sigma^{total} - d\sigma^{BH})}{\int_{0}^{2\pi 
}d\phi(d\sigma^{total} - d\sigma^{BH})}. 
\label{AAA} 
\end{equation} 
$\mathcal{A}$ describes the asymmetry for the proton and the positron to be found in the same and 
opposite hemispheres. 
It is non-zero only due to the interference term. 
As shown in \cite{asymmetry, asymmetry1} the asymmetry $\mathcal{A}$ 
gives an access to the real part of the DVCS amplitude, which in turn allows 
to extract SPD. 
 
In case of a polarized positron beam and unpolarized target the contribution to the total cross section coming from the interference term 
can be written in leading order (using the notation of \cite{diehl}) as: 
\begin{eqnarray} 
(\tau_{BH}^*\tau_{DVCS} + \tau_{DVCS}^*\tau_{BH})_{pol} = \nonumber \\ 
 \frac{4\sqrt{2} m e^6}{tQx} \cdot \frac{1}{\sqrt{1-x}} \cdot 
e_lP_l\left[-\sin\phi\cdot\sqrt{\frac{1+\epsilon}{\epsilon}}Im\tilde{\mathcal{M}}^{1,1}\right], 
\label{phi} 
\end{eqnarray} 
where $\tau_{BH}$ and $\tau_{DVCS}$ are the BH and DVCS amplitudes, $\tilde{\mathcal{M}}^{1,1}$ is the linear combination of DVCS helicity amplitudes that contributes in the polarized 
case, $\epsilon$ is the polarization parameter of the virtual photon while $e_l$ and $P_l$ denote lepton charge and 
polarization of the incident lepton, respectively. 
As it was already mentioned -- in contrast to the pure BH or DVCS 
contributions, where the real and imaginary parts of the amplitude are mixed up and 
difficult to disentangle -- 
the determination of the $\sin\phi$-moment of the asymmetry of the interference term with respect to the beam 
polarization provides information on the imaginary part of $\tilde{\mathcal{M}}^{1,1}$, 
which is directly related to SPD \cite{diehl}. 
%
 
\subsection{Monte Carlo simulations} 
ZEUS and H1 have each written Monte Carlo (MC) generators based on the calculations of Frankfurt, Freund and Strikman (FFS) \cite{ffs}, to simulate 
the elastic DVCS 
and BH processes and interference between them. 
Also Donnachie and Dosch (DD) \cite{dodo} published their calculations of the DVCS cross section. 
Both these predictions provide the scattering amplitude at $t = t_{min} \simeq -m_p^2 Q^4 / W^4$, where $t$ is the 
squared momentum transfer to the proton, $t_{min}$ its minimum value, $m_p$ the proton mass and $W$ the invariant mass of the $\gamma ^* 
p$ system. An exponential $t$-dependence, $e^{-b\mid t\mid}$, is assumed. 
 
 
\section{Event selection} 
\subsection{Event signatures in detector} 
The DVCS and BH events have a very simple signature in the detector. Since the proton escapes down the beam-pipe 
only the positron and the photon can be seen. In case of BH the photon is emitted from the positron lines, so 
the highest probability is to find both the positron and the photon in the backward\footnote{the outgoing proton beam defines the forward direction} part of the detector.  
The DVCS process has a different nature, so the ratio of DVCS over BH events is expected to increase 
when the photon is found in the central/forward direction. 
The selection criteria are chosen in such a way that the detector acceptance is high and the expected 
contribution of DVCS to the total cross section is of the same order as that of BH. 
 The products of the DVCS process are seen in the detector as two electromagnetic clusters : the positron emitted into a 
backward detector and the photon found in the central/forward calorimeter. For most of these events no track is 
reconstructed  
due to the limited acceptance of the backward tracking devices.  In the BH case, events are selected with a signature identical to that of 
the DVCS process but, in addition, events where the photon is emitted backwards and the positron is 
found in the central/forward calorimeter. These are characterized by a track linked to the  
electromagnetic cluster in the central/forward calorimeter.  
  
 
 
\subsection{Selection cuts} 
The details of the selection criteria differ slightly for the ZEUS and H1 cases, however the general idea stays 
the same. 
Selected are events with: 
\begin{itemize}
 \parskip=0pt
   \parsep=0pt
   \itemsep=0pt 
\item two electromagnetic clusters: a high energetic one detected in the backward calo\-ri\-meter and one
with transverse momentum $>$ 1 GeV found in central/forward calorimeter, 
\item lack of any other activity above the noise threshold in the calorimeter and empty forward detectors -- 
to eliminate dissociative events, 
\item no more than one track reconstructed; if the track is found, it has to be linked to one of the 
clusters -- the cluster with the track is identified as the positron; when no track is found the backward 
cluster is assumed to be the positron, 
\item $Q^2$ bigger then a few GeV -- to justify the use of perturbative QCD in theoretical predictions. 
\end{itemize} 
 
\section{Analysis, Results and Discussion} 
\subsection{ZEUS -- the first observation of DVCS} 
\label{zeus_old}
The results of the first observation of DVCS at HERA were reported by the ZEUS  
Collaboration \cite{zeus}.  
The data used for the DVCS analysis were collected during 1996-97 and  
correspond to an integrated luminosity 
of  37 $\mbox{\rm pb}^{-1}$. 
 
\begin{figure}[h] 
\begin{center} 
\mbox{ 
\epsfig{file=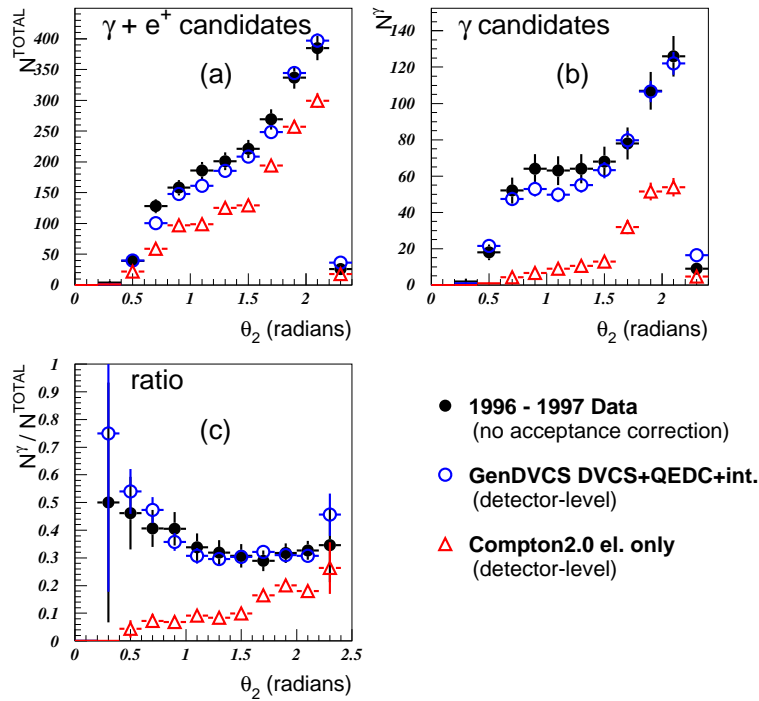, 
height=10.cm, 
clip=} 
} 
\end{center} 
\vspace*{-0.5cm}
\caption{\label{zeus_1}  
Distributions of the 
polar angle of electromagnetic cluster found in the central/forward calorimeter. (a) - all selected 
events, (b) - DVCS candidates only, (c) - ratio of (b) to (a). The uncorrected data (solid points) are 
compared to the BH (empty triangles) and DVCS+BH+interference of these two (open circles)  
predictions.} 
 
\end{figure} 
 
The selected events are plotted - Figure~\ref{zeus_1}a - as a function of the polar angle of the 
electromagnetic cluster found in the central/forward calorimeter. 
The data are compared to the MC predictions for the BH process 
as generated by 
Compton2.0 \cite{compton2} and to DVCS + BH + interference as predicted by the DVCS MC. 
All MC predictions plotted in Figure~\ref{zeus_1} are normalized to the same  
luminosity as the data. 
 
Figure~\ref{zeus_1}b shows similar distributions but only for DVCS candidates, i.e. the events where the electromagnetic 
cluster in the central/forward calorimeter is identified as a photon (no track is linked). 
Both plots (Figure~\ref{zeus_1}a,b) indicate that the BH process alone is not able to describe the data and only the 
inclusion of the DVCS part brings MC into a reasonable agreement with the data. 
 
 
Although the selection procedure is tailored to eliminate the inelastic events, still some 
contribution (expected to be of the order of 20\%) remains in the selected sample. Dissociative 
events are not present in any of the MCs used for the analysis, so one has to keep in 
mind that the MC predictions have to be raised by roughly this amount in Figure~\ref{zeus_1}a,b. 
A distribution that is found to be insensitive to the inelastic contribution is the ratio of 
DVCS candidates to all selected events, plotted in Figure~\ref{zeus_1}c. In addition, the efficiency of finding 
electromagnetic cluster cancels for this distribution. 
It can be noticed that the conclusions drawn on the basis of two previous distributions hold also in this case. It is now clearly seen that  
especially for small angle photons there is a clear deficit of events in the 
BH prediction.

\begin{figure}[h] 
\begin{center} 
\mbox{ 
\epsfig{file=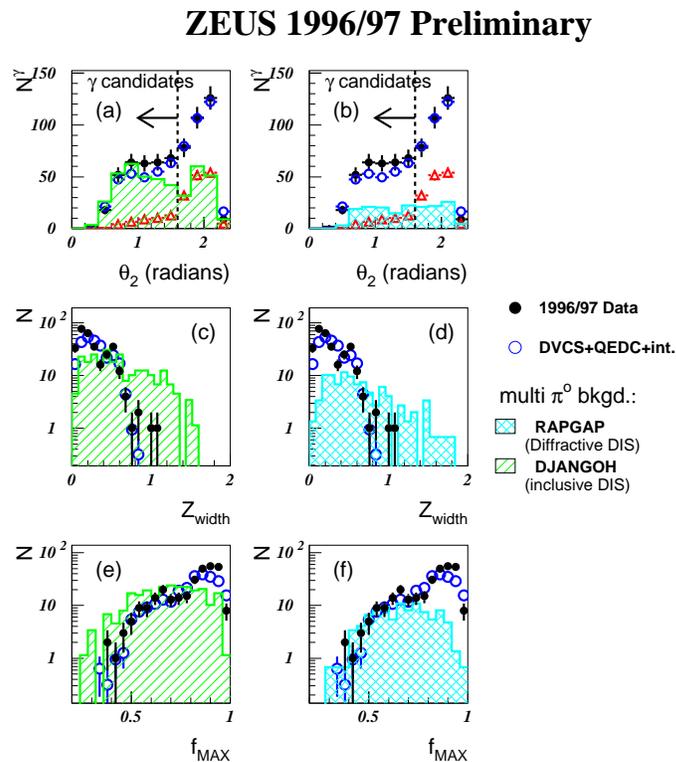, 
height=10.cm, 
clip=} 
} 
\end{center} 
\vspace*{-0.5cm}
\caption{\label{zeus_3}  
Distribution of (a,b) -- the 
polar angle of electromagnetic cluster found in the central/forward calorimeter, 
(c,d) -- the energy weighted z-position of the electromagnetic cluster  
expressed in units of the electromagnetic cell width, 
(e,f) -- the fraction of the electromagnetic cluster energy carried by the most energetic cell in the cluster.   
} 
\end{figure}

 
A potential source of background arises from $\pi ^0/ \eta$ production with the decay photons reconstructed in a single cluster. To investigate this background  
once more plots of the polar angle of the electromagnetic cluster found in the  central/forward calorimeter are made, but this time also the predictions from  DJANGOH (Figure~\ref{zeus_3}a) and RAPGAP (Figure~\ref{zeus_3}b) are overlayed. 
Both these MCs are expected to provide a hadronic background according to the reactions: 
$e^+p \rightarrow e^+ p\, \pi ^0 \pi ^0, e^+p \rightarrow e^+ p\,\pi ^0 \eta$ etc. 
It can be noticed that DJANGOH predictions are similar in shape but about twice as large as RAPGAP ones. 
 
It should be stressed that both the generators (DJANGOH, RAPGAP) 
are high multiplicity  
 MCs and cannot be expected to predict accurate rates for the single $\pi^0/ \eta$ 
production. Moreover, calculations of rates expected at HERA, based on low  
energy data, show that  
in the kinematic region where the measurement is performed one cannot expect 
more than a few $\pi ^0/ \eta$.  
Therefore, the predictions of the high multiplicity MCs seem to largely 
overestimate the single $\pi ^0/ \eta$ background in DIS and cannot be 
relied on.  
 
Another way to study the possibility of the $\pi ^0/ \eta$ background is the analysis of the shower shapes. It can be expected that the $\pi ^0/ \eta$ clusters --  
since built by two particles -- should be broader and larger, and the deposit of energy in a single calorimeter cell ought to be smaller than in case of a single photon cluster. 
For the purpose of this study two shower shape variables are defined: 
\begin{itemize}
\parskip=0pt
   \parsep=0pt
   \itemsep=0pt 
 \item energy weighted  average of the width of the cluster in the z--direction ($z_{width}$) 
\begin{equation} 
z_{width} = \frac{\Sigma (\mid z_{cell} - \overline{z}\mid \cdot E_{cell})} {\Sigma E_{cell}}, 
\end{equation} 
where the sum is over all cells in the electromagnetic cluster,  
\item the fraction of the electromagnetic cluster energy which is deposited in  
the most energetic cell in the cluster $f_{max}$ 
\begin{equation} 
f_{max} = \frac{energy\;\; of\;\; the\;\; most\;\; energetic\;\; cell\;\; in\;\; the\;\; cluster}{total\;\; energy\;\; in\;\; the\;\; cluster}. 
\end{equation} 
\end{itemize} 
The distributions of the selected ZEUS data as a function of these two shower shape variables are shown in Figure~\ref{zeus_3}c-f and compared to the $\pi ^0 / \eta$ shower shapes as generated by DJANGOH and RAPGAP. 
These plots point out that the clusters reconstructed in the data have the same shapes as the photon clusters generated by DVCS 
MC. At the same time the $\pi ^0/ \eta $ showers produced by DJANGOH and RAPGAP seem to be quite different since 
they have too small $f_{max}$ and too large $z_{width}$. 
 
The results indicate that the clusters seen in the data have different origins then those produced by $\pi ^0/ \eta$. 
Thus, the hadronic background from low multiplicity processes 
cannot account for the data excess above the BH prediction. 
 
\subsection{H1 -- the first measurement of the DVCS cross section} 
The H1 Collaboration, made one step further and measured the DVCS cross section \cite{h1}. 
 For this analysis H1 used the data collected in 1997 running period which corresponds to an integrated luminosity of $ 8~\mbox{\rm pb}^{-1}$.
 
The selected data were divided into two samples: 
\begin{itemize} 
\parskip=0pt
   \parsep=0pt
   \itemsep=0pt 
\item control sample -- characterized by the photon candidate detected in the backward calorimeter and the positron candidate in the central/forward part. This sample is dominated by the BH contribution. 
\item enriched DVCS sample -- characterized by the positron candidate in the backward calorimeter 
and the photon in the central one. Both DVCS and BH contribute to this sample. 
\end{itemize}\begin{figure}[h] 
\vspace*{-0.3cm}
\begin{center} 
\mbox{ 
\epsfig{file=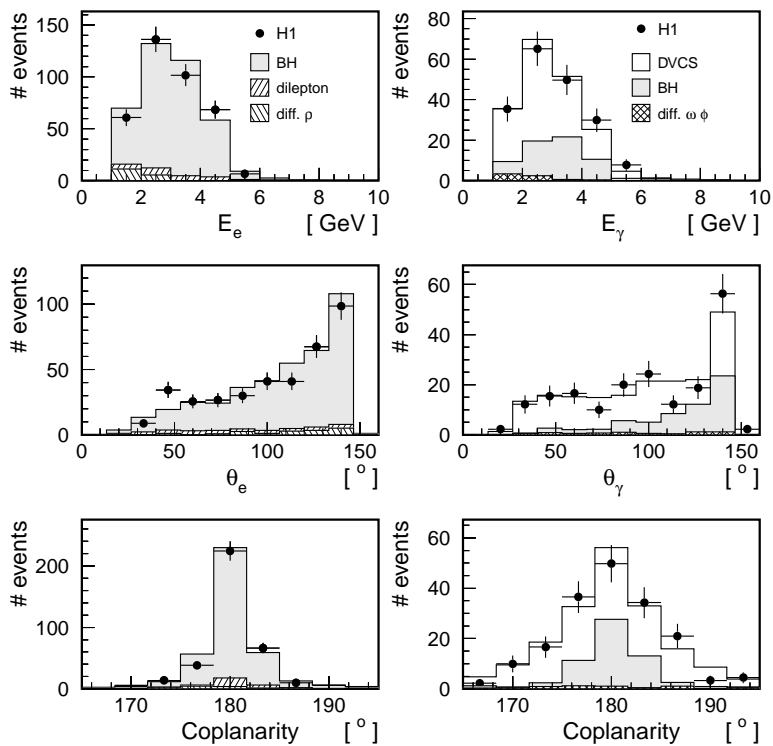, 
height=10.cm, 
bbllx=6pt,bblly=155pt, 
bburx=600pt,bbury=660pt, 
clip=} 
} 
\end{center} 
\vspace*{-0.4cm}
\caption{\label{h1_1}  
Distribution of (a,b) -- energy of the cluster found in the central/forward calorimeter, 
(c,d) -- polar angle of the cluster reconstructed in the central/forward calorimeter, 
(e,f) -- coplanarity i.e. the difference of the azimuthal angle of the positron and photon candidates.  Left column -- events from the control sample, right column -- data from the enriched DVCS sample. 
The data (solid points) are compared to the sum of predictions of different MC models. Plots in the left column are 
normalized to luminosity whereas those in the right column 
are normalized to the total number of events (i.e. all MCs but DVCS are normalized to the luminosity, and the total sum of events from different MC predictions is equal to the number of events from enriched DVCS sample). 
} 
\vspace*{-0.5cm}
\end{figure} 
 
The cross section measurement is based on the enriched DVCS sample and the control part is used as a reference sample 
to monitor the detector performance and its simulation. In order to have control of the detector response in the same 
energy and angular ranges as for the enriched DVCS sample, for the calculation of kinematic variables the cluster in the central/forward calorimeter is always treated as the photon and the one in the backward calorimeter as the positron. 
The control sample is populated mainly by elastic BH events. However, it also contains small contributions from 
inelastic BH (estimated to be $7.7 \pm 3.8 \%$) as well as some events from diffractive 
electroproduction of $\rho $ mesons ($\rho 
\rightarrow \pi ^+ \pi^-$) and the elastic production of electron pairs ($e^+p \rightarrow e^+e^-e^+p$).    
Due to the large scattering angle of the positron, the DVCS process in this sample is suppressed to negligible levels. 
In the left column of Figure~\ref{h1_1} the data from the control sample are plotted and compared to the sum of MC 
predictions of the BH process, elastic $\rho$ production and elastic dilepton production, all normalized to luminosity. 
It may be noticed that the sum of MCs provide good overall description of the data, showing that the detector response is under control and well described by the simulation. 
 
Although the main contribution to the enriched DVCS sample comes from the elastic DVCS and BH processes also different 
background sources have to be considered. The contamination of inelastic DVCS and BH events has been estimated to be 
$16 \pm 8 \%$ of the final sample. 
Other background sources are due to the diffractive $\omega $ ($\omega \rightarrow \pi^0 \gamma$) and $\phi $ 
($\phi \rightarrow K_l^0K_S^0,\; K_S^0 \rightarrow \pi^0 \pi^0$) production and are estimated to be $3.5 \%$. 
The background arising from $\pi^0$ production in low 
multiplicity DIS, with the decay photons reconstructed in a single cluster, is estimated from the data and found to be 
negligible. 
 
In the right column of Figure~\ref{h1_1} various event distributions for the data from enriched DVCS sample are compared 
to the sum of MCs predicting contributions of all relevant processes. 
All MCs but DVCS are normalized to luminosity of the data. 
The DVCS part is normalized in such a way that the sum of contributions from different MCs is equal to the 
total number of events in the data. 
It is worth noticing that the BH prediction not only fails to describe the normalization of the data but also predicts a  
different shape for some distributions. In particular, the differences in shape are seen for the  polar angle and 
coplanarity, where the latter is defined as the difference in azimuthal angle of the photon and positron clusters. 
The coplanarity for the data is much broader than the one predicted by the BH MC. 
This is attributed to the electromagnetic nature of the BH process which has a steeper $t$--dependence than
the DVCS signal. 
 
\begin{figure}[h] 
\vspace*{-0.5cm} 
\begin{center}  
\mbox{ 
\epsfig{file=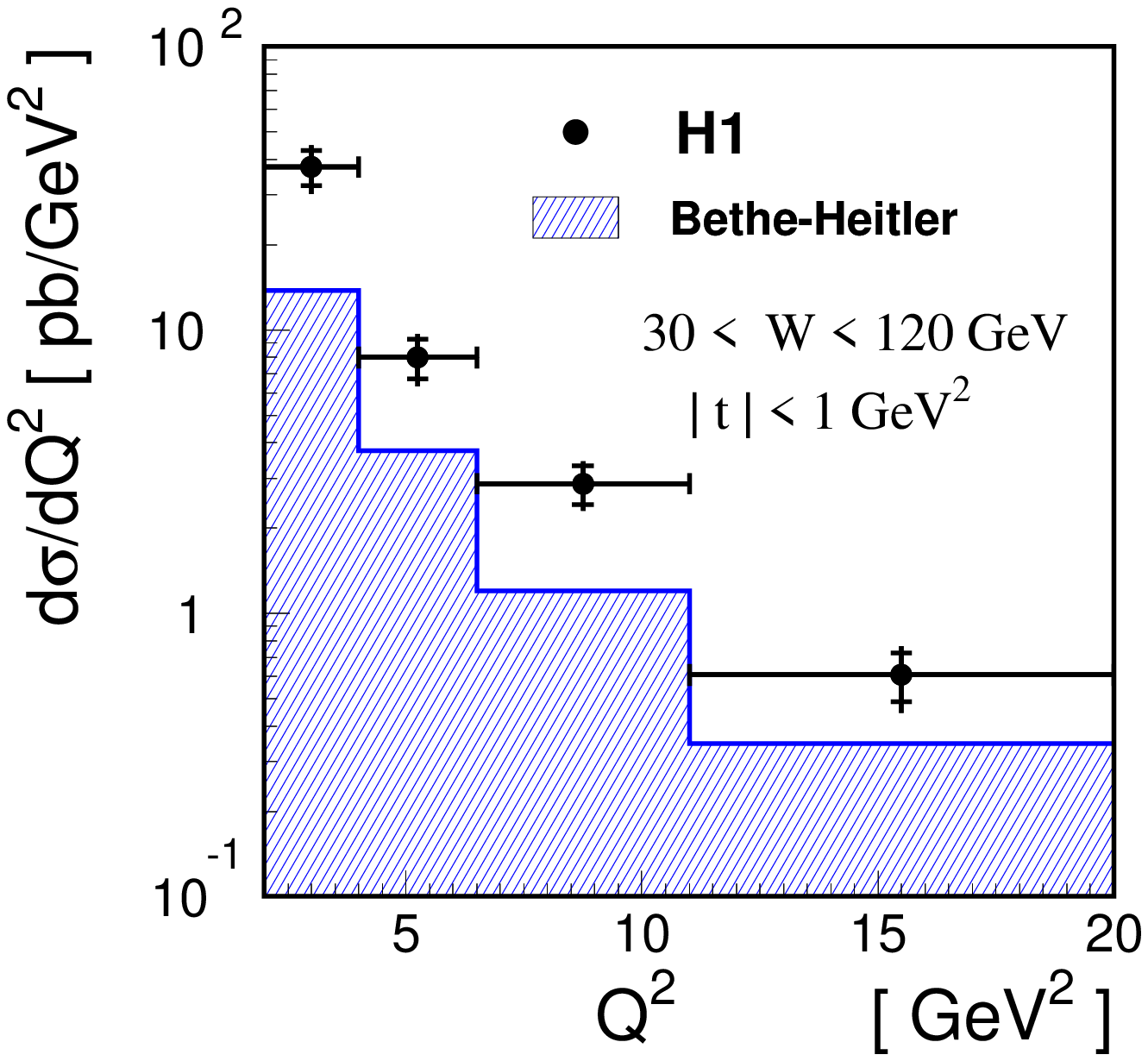, 
height=6.5cm, 
bbllx=80pt,bblly=235pt, 
bburx=500pt,bbury=630pt, 
clip=} 
\epsfig{file=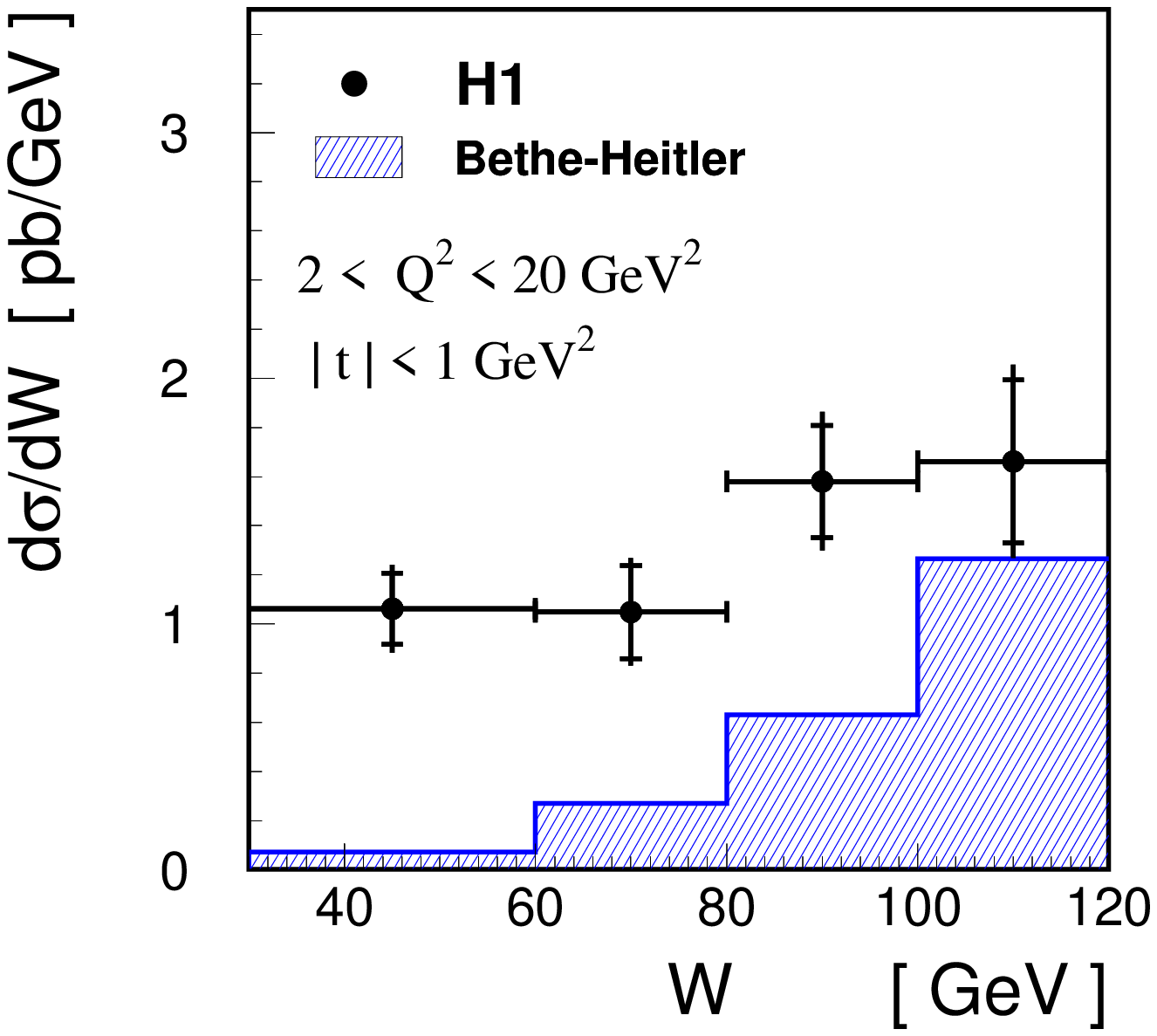, 
height=6.5cm, 
bbllx=80pt,bblly=235pt, 
bburx=900pt,bbury=630pt, 
clip=} 
} 
\vspace*{-0.5cm}
\caption{\label{h1_2}  
Differential cross section for the reaction $e^+p \rightarrow e^+p\gamma $ as a function of $Q^2$ (a) and $W$ (b). The data (solid points) are plotted with statistical (inner error bar) and systematic errors added in quadrature. The hatched histogram shows the contribution of BH process.  
} 
\end{center}
\vspace*{-0.5cm} 
\end{figure}

To extract the cross section the data have been corrected for acceptance, detector effects and  
initial state radiation (radiation of a real photon from the positron line). Also various background contributions have 
been subtracted. 
In Figure~\ref{h1_2} the cross section is presented differentially in $Q^2$ and $W$. The measurement is performed in the 
kinematic region defined by: $2 < Q^2 < 20\; GeV^2$, $30 < W < 120\;GeV$, $\mid t \mid < 1\;GeV^2$. The data are 
compared to the BH prediction. It is noticed that at small $W$ values, the total cross section is dominated by 
the DVCS contribution, while for large $W$, BH is dominant. Unfortunately, the limited resolution and statistics do not allow to measure the cross section differentially in $t$ and to extract the $t$-slope.
Data points in Figure~\ref{h1_2} are plotted with statistical and systematical errors added in quadrature.  
The total systematic error is found to be around $15\%$. 
The main contribution to this error ($8\%$) is due to the uncertainty of the measurement of the angle of the scattered positron, because for most of the events no vertex can be reconstructed.  
Another significant contribution ($\sim 8\%$) comes from the estimate of the  
contamination of inelastic events. 
\begin{figure}[h] 
\vspace*{-0.5cm} 
\begin{center} 
\mbox{ 
\epsfig{file=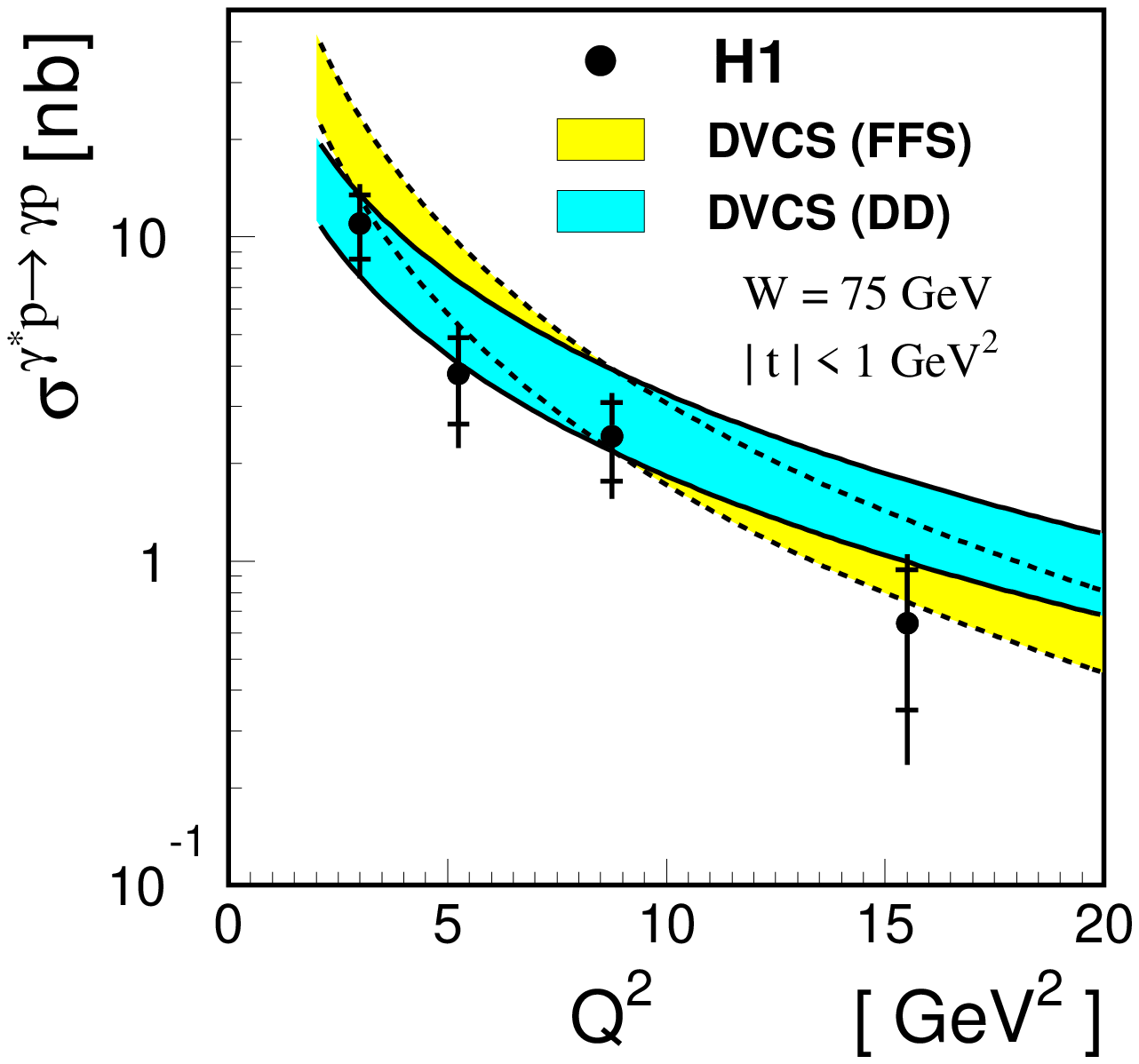, 
height=6.5cm, 
bbllx=80pt,bblly=235pt, 
bburx=500pt,bbury=630pt, 
clip=} 
\epsfig{file=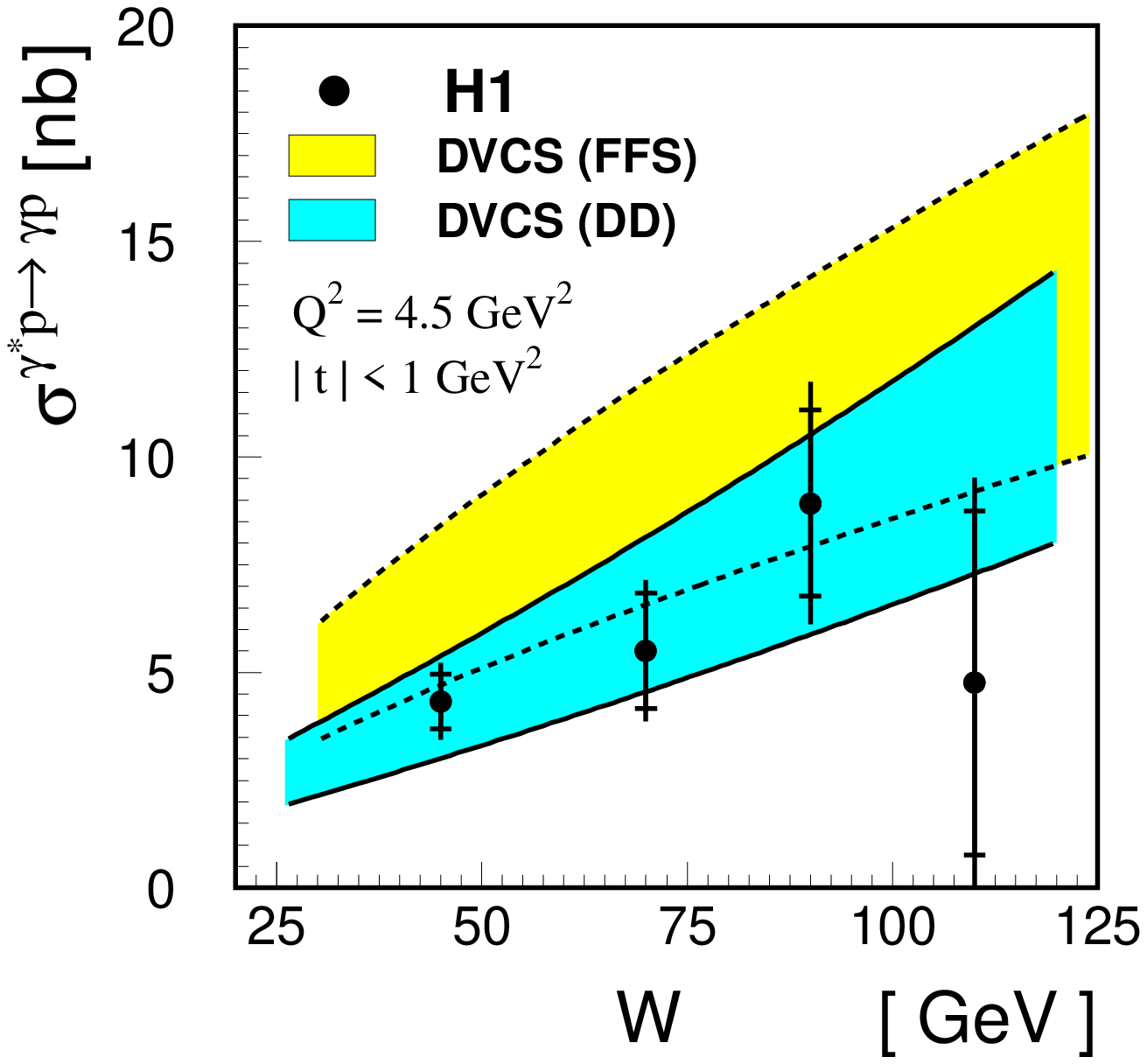, 
height=6.5cm, 
bbllx=80pt,bblly=235pt, 
bburx=900pt,bbury=630pt, 
clip=} 
} 
\end{center} 
\vspace*{-0.3cm}
\caption{\label{h1_3}  
$\gamma^*p \rightarrow \gamma p$ DVCS cross section as a function of $Q^2$ (a) and $W$ (b). 
The data (solid points) are plotted with statistical (inner error bar) and systematic errors added in quadrature. 
Theoretical predictions are shown with gray band. The band width comes from the theoretical uncertainty connected 
with the $t$-slope which is assumed to be between 5 (upper edge of the band) and 9 $\mbox{\rm GeV}^{-2}$ (lower edge). 
} 
\end{figure} 
On the basis of the explanation given in section \ref{discussion},  
the DVCS cross section is extracted from the total one by subtracting the BH contribution. 
The result is then converted to the $\gamma^*p \rightarrow \gamma p$ DVCS cross section, plotted in Figure~\ref{h1_3}. 
The theoretical predictions of FFS \cite{ffs} and DD \cite{dodo} are also overlayed. However, due to the unknown 
$t$-slope the absolute normalization of the theoretical predictions is uncertain. The assumption of the $t$-slope 
being in the range between 5 to 9 $\mbox{\rm GeV}^{-2}$ -- as suggested by the light vector meson measurements -- 
results in the theoretical predictions seen as the bands, which lower edges correspond to the higher limit of 
the $t$-slope and the upper edges to the lower $t$-slope bound. 

The data are, within errors, in agreement with both theoretical models. 

\subsection{ZEUS -- the cross section measurement}
Recently also ZEUS presented the results of the DVCS cross section measurement \cite{zeus_new}. 
As in the previous analysis the selected events were divided into two samples:
one characterized by a positron in the central calorimeter (positron sample) and the other one with a photon 
in that part of the detector (photon sample). The positron sample, after subtraction of a small contribution
from di-electron events, was found to be in excellent agreement with BH MC predictions. The BH background
 was then subtracted from the photon sample using the BH MC prediction normalized according to the positron
sample. 
Finally, the data were corrected for detector smearing and acceptance and the systematic uncertainties were 
analysed. The major contributions to systematic uncertainties come from the BH MC description, uncertainty in determining the hadronic background and the energy scale uncertainty.

\begin{figure}[h] 
\vspace*{-0.5cm} 
\begin{center}  
\mbox{ 
\epsfig{file=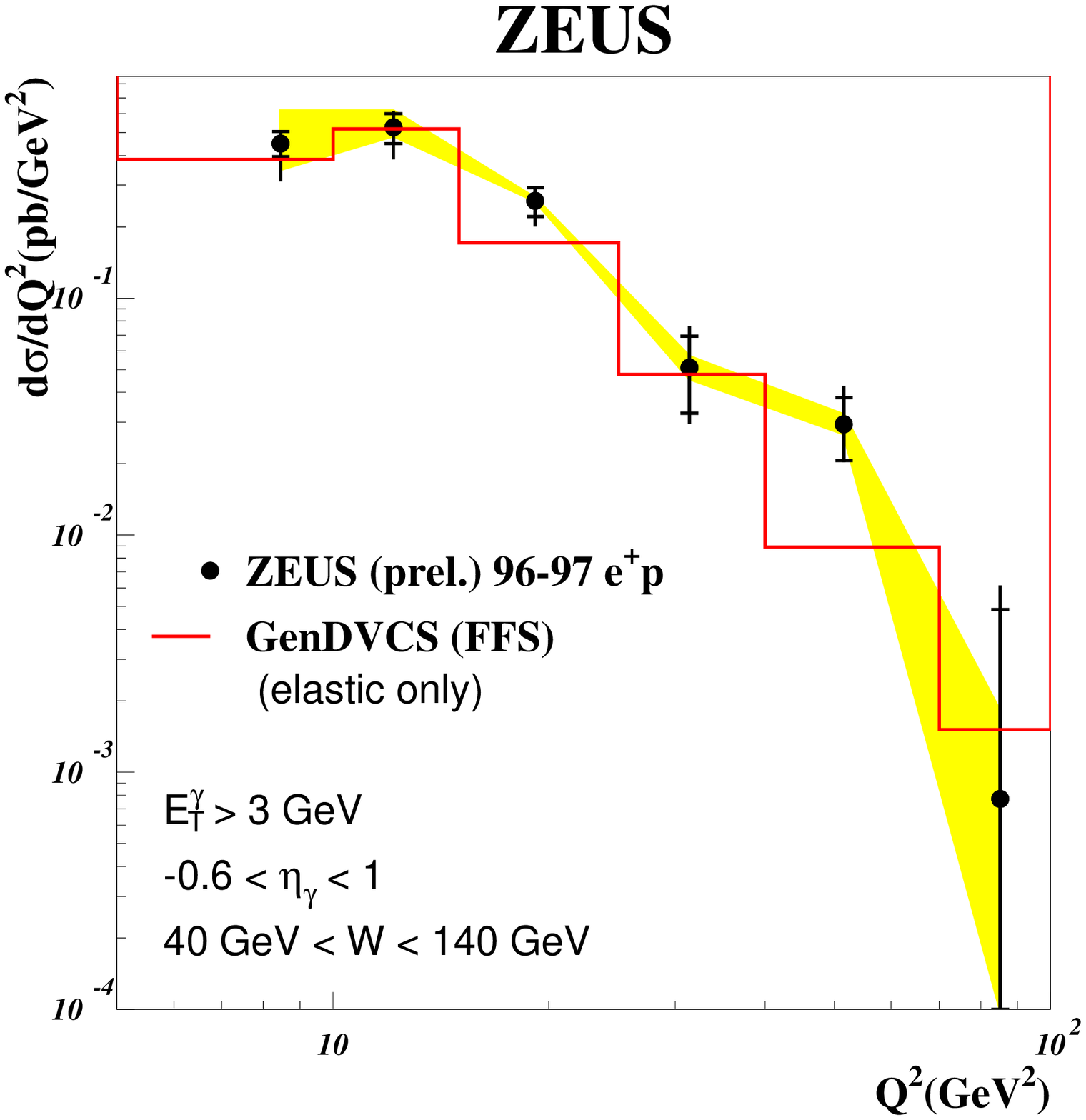, 
height=7.2cm, 
bbllx=10pt,bblly=135pt, 
bburx=700pt,bbury=730pt, 
clip=} 
\epsfig{file=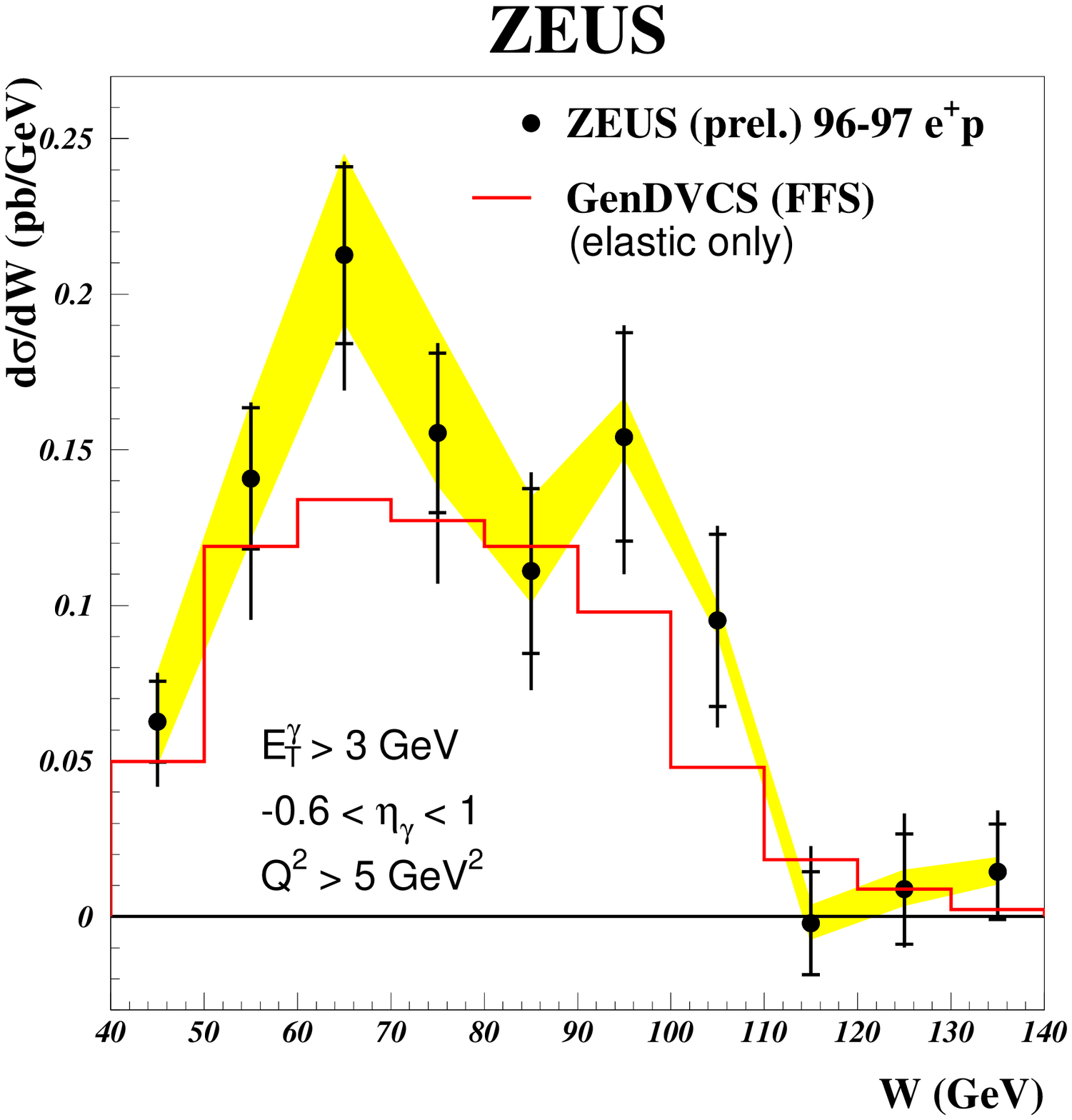, 
height=7.2cm, 
bbllx=10pt,bblly=135pt, 
bburx=900pt,bbury=730pt, 
clip=} 
} 
\vspace*{-0.5cm}
\caption{\label{zeus_new}  
Differential DVCS cross section as a function of $Q^2$ (left) and $W$ (right). The data (solid points) are plotted with statistical (inner error bar) and systematic errors added in quadrature. The calorimeter energy scale uncertainty, which is correlated between bins, is shown separately as the shaded band. The histogram shows the DVCS MC prediction.  
} 
\end{center}
\vspace*{-0.5cm} 
\end{figure} 

The cross section is measured in the kinematic region defined by $Q^2 > 5\; GeV^2$, $40 < W < 140\;GeV$, 
$E_T^{\gamma}> 3\;GeV$ and $-0.6 < \eta_{\gamma} < 1.0$, where $E_T^{\gamma}$ and $\eta_{\gamma}$ are the 
transverse energy and pseudorapidity of the final state photon, respectively. 
In Figure~\ref{zeus_new} the DVCS cross section is shown as a function of $Q^2$ and $W$. The data are plotted 
with statistical errors and systematic uncertainties added in quadrature. The energy scale uncertainty, 
which is correlated between bins, is shown as a band.  The data are compared with the predictions of the DVCS MC. The predictions are in general agreement with the data, both in shape and normalization. 
However, as mentioned in section \ref{zeus_old} the data include also a small (about 20\%) contribution of inelastic events, so the MC predictions should be risen by roughly this amount. It can be noticed that the correction for dissociative events should improve the overall agreement between the data and DVCS MC.


\subsection{HERMES -- the beam-spin asymmetry in hard exclusive electroproduction of photons.}  
\begin{figure}[h] 
\begin{center} 
\mbox{ 
\epsfig{file=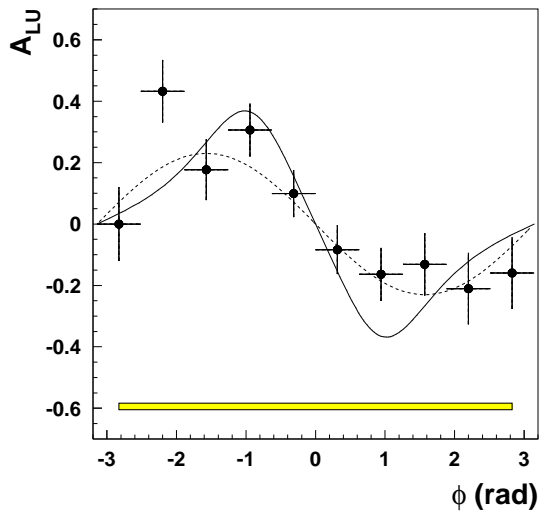, 
height=7.cm, 
bbllx=60pt,bblly=105pt, 
bburx=600pt,bbury=610pt, 
clip=} 
} 
\end{center} 
\vspace*{-0.2cm}
\caption{\label{hermes_1}  
Beam-spin asymmetry from HERMES as a function of $\phi$ for the missing mass range $-1.5 < M_x < 1.7$. 
The dashed line shows $0.23\cdot \sin\phi  $ function and the solid one the curve 
calculated taking into account SPD \cite{polyakov}. \\ 
The systematic uncertainty is represented by the error band shown at the bottom of the figure. 
} 
\end{figure}

The picture of DVCS measurements at HERA cannot be complete without the recently published HERMES results \cite{hermes} 
of the beam-spin asymmetry analysis. 
The data used by HERMES were collected in the years 1996-97. A longitudinally polarized positron beam and a hydrogen target were used. 
In contrast to the ZEUS and H1 measurements, HERMES studies observables directly connected to the interference term 
between the DVCS and BH processes. 
In Figure~\ref{hermes_1}, the $\phi$-dependence of the beam-spin asymmetry $A_{LU}$ is plotted, 
\begin{equation} 
A_{LU}(\phi )=\frac{1}{\langle |P_l| \rangle} \cdot \frac{N^+(\phi) - N^-(\phi )}{N^+(\phi) + N^-(\phi )}, 
\end{equation} 
where $N^+$ and $N^-$ stand for the luminosity normalized yields of events with corresponding beam helicity states, 
$ \langle |P_l| \rangle $ means the average magnitude of the beam polarization, and the subscripts $U$ and $L$ 
denote unpolarized target and longitudinally polarized beam, respectively. 
Events contributing to this plot are required to have missing mass $M_x$  between -1.5 and 1.7 $\mbox{\rm GeV}$, i.e. in the range $-3\sigma $ to $+1\sigma $ around the proton mass. The missing mass is defined as $M_x^2 = (q+P_p-k)^2$, where $q$, $P_p$ and $k$ denote the four momenta of the virtual photon, 
the target nucleon and the real photon, 
respectively\footnote{Due to the limited momentum resolution 
$M_x^2$ may be negative and then $M_x=-\sqrt{-M_x^2}$ is defined.}. 
 The limits for the missing mass $M_x$ required by this measurement are chosen 
 in such an asymmetric way, in order to minimize the influence of the DIS-fragmentation background 
 while optimizing the statistics. 
The data are compared to a simple $\sin\phi $ curve and to the model of Ref.~\cite{polyakov} 
that takes into account SPD. The agreement between the data points and $\sin\phi$ function demonstrates that 
the $\phi$-dependence is consistent with the expectations of equation~(\ref{phi}). 
In addition, the $\sin\phi$-weighted moments are defined: 
\begin{equation} 
A_{LU}^{\sin \phi^\pm} = \frac{2}{N^\pm}\sum_{i=1}^{N}\frac{\sin\phi_i}{|P_l|_i},  
\end{equation} 
and used to analyze the beam-spin asymmetry for different missing mass bins. It turns out that the beam-spin asymmetry 
vanishes for higher missing masses ($M_x > 1.7 \mbox {\rm GeV}$), and that the sign of the $\sin\phi$ moment is opposite for the two 
beam helicities -- which is in agreement with the expectations for the helicity dependence of the relevant DVCS-BH interference term. 
 
\section{Conclusions and Prospects} 
The HERA results  
 constitute the first step in studies of the DVCS process itself, as well as in extraction of SPD 
by means of analysis of DVCS and its interference with BH. 
It is obvious that these studies do not answer all questions and do not fulfill all expectations connected with the 
measurement. 
However, the vivid theoretical interest in this process 
\cite{polyakov,new_theory} along with the HERA upgrade resulting in higher luminosity and the detector 
modifications (e.g. at H1 the improved performance of backward tracker and installation of the very forward proton 
spectrometer will allow for a direct $t$-measurement and an elastic/dissociative proton separation ) 
give hope that at the next Ringberg workshop much more 
information  
regarding DVCS and SPD will be presented. 
In particular, the $t$-dependence of the DVCS cross section and azimuthal angle 
asymmetries are planned to be measured by H1 and ZEUS.
 

\ack
The author would like to thank the organizers for the kind invitation and perfect organization of the 
Ringberg workshop. 
 
     

\Bibliography{xx} 
\bibitem{spd} M\"{u}ller D et. al. 1994  {\it Fortsch. Phys.} {\bf 42} 101, Ji X 1997 {\it Phys. Rev. Lett. } {\bf 78} 610, Radyushkin A 1997 {\it Phys. Rev.} D {\bf 56} 5524, 
Golec-Biernat K and Martin A 1999 {\it Phys. Rev.} D {\bf 59} 014029 
\bibitem{factorization}Collins J C and Freund A 1999 {\it Phys. Rev.} D {\bf 59} 074009 
\bibitem{dvcs_theory} Ji X 1997 {\it Phys. Rev.} D {\bf 59} 074009, Ji X and Osborne J 1998 {\it Phys. Rev.} D {\bf 58} 094018, Bl\"umlein~J and Robaschnik~D 2000 {\it Nucl. Phys. } B {\bf 581} 449 
\bibitem{asymmetry} Freund A 2000 {\it Phys. Lett.} B {\bf 472} 412 
\bibitem{asymmetry1} Freund A and McDermott M 2001 {\tt hep-ph/0106124}  
\bibitem{diehl} Diehl M et. al., 1997 
              {\it Phys. Lett.} B {\bf 411} 193 
\bibitem{polyakov} Kivel N, Polyakov M and Vanderhaeghen M 2001 {\it Phys. Rev.} D {\bf 63} 114014 
\bibitem{new_theory} McDermott M 2001 {\tt hep-ph/0107224}, Freund~A and McDermott M 2001 {\tt hep-ph/0106115}, Freund~A and McDermott M 2001 {\tt hep-ph/0106319}, Korotkov V A and Nowak W D 2001 {\tt hep-ph/0108077}  
\bibitem{zeus} Saull P R B 2000 {\tt hep-ex/0003030} 
\bibitem{h1} Adloff C [H1 Collaboration] 2001 {\it Phys.Lett.} B {\bf 517} 47 \bibitem{zeus_new} [ZEUS Collaboration] 2001 {\it paper submitted to International Europhysiscs Conference on High Energy Physics in Budapest}
\bibitem{hermes} Airapetian A [HERMES Collaboration] 2001 {\it Phys. Rev. Lett.} {\bf 87} 182001  
\bibitem{clas} Stepanyan S [CLASS Collaboration] 2001 {\it Phys. Rev. Lett.} {\bf 87} 182002
\bibitem{compton2} Courau A, Kermiche S, Carli T and Kessler P 1991  {\it Proc.\
of the Workshop on Physics at HERA (Hamburg)} vol~2  p~902 
\bibitem{ffs} Frankfurt L L, Freund A and Strikman M 1998 {\it Phys.~Rev.} D {\bf 58} 114001 and 1999 {\it erratum} {\it Phys.~Rev.} D {\bf 59} 119901E 
\bibitem{dodo} Donnachie A and Dosch H G 2001 {\it Phys. Lett.} B {\bf 502} 74 
 
 
 
\endbib
\end{document}